\documentclass[11pt,a4paper]{article}

\usepackage{bm,bbm,amsfonts,amssymb,mathrsfs,amsmath,ascmac}
\usepackage{graphicx}
\usepackage{hyperref}
\usepackage{color}

\newcommand{\Z}{\mathbb{Z}}

\newcommand{\bra}[1]{\langle #1 \rvert}
\newcommand{\ket}[1]{\lvert #1 \rangle}

\newcommand{\Bket}[1]{\lvert #1 \rangle\!\rangle}
\newcommand{\ol}{\overline}
\newcommand{\wt}{\widetilde}
\newcommand{\wh}{\widehat}
\newcommand{\cR}{\mathcal{R}}
\newcommand{\cT}{\mathcal{T}}
\newcommand{\cA}{\mathcal{A}}
\newcommand{\cC}{\mathcal{C}}
\newcommand{\cJ}{\mathcal{J}}

\newcommand{\tot}{{\rm tot}}

\makeatletter

\@addtoreset{equation}{section}
\makeatother

\date{\today}

\begin{document}

\begin{titlepage}

\renewcommand{\thefootnote}{\fnsymbol{footnote}}

\begin{flushright}
 {\tt 
 IPHT-T14/014\\
 RIKEN-MP-84
 }
\\
\end{flushright}

\vskip9em

\begin{center}
 {\Large {\bf 
 Current Reflection and Transmission at Conformal Defects: 
 Applying BCFT to Transport Process
 }}

 \vskip5em

 \setcounter{footnote}{1}
 {\sc Taro Kimura$^{(a,b)}$}\footnote{E-mail address: 
 \href{mailto:taro.kimura@cea.fr}
 {\tt taro.kimura@cea.fr}} 
 and
 \setcounter{footnote}{2}
 {\sc Masaki Murata$^{(c)}$}\footnote{E-mail address: 
 \href{mailto:m.murata1982@gmail.com}
 {\tt m.murata1982@gmail.com}}

 \vskip2em

{\it 
$^{(a)}$Institut de Physique Th\'eorique,
 CEA Saclay, F-91191 Gif-sur-Yvette, France
 \\ \vspace{.5em}
 $^{(b)}$Mathematical Physics Laboratory, RIKEN Nishina Center, 
 Saitama 351-0198, Japan 
 \\ \vspace{.5em}
 $^{(c)}$Institute of Physics AS CR, Na Slovance 2, Prague 8, Czech Republic
}

 \vskip3em

\end{center}

 \vskip2em

\begin{abstract}

We study reflection/transmission process at conformal defects by
 introducing new transport coefficients for conserved currents.
These coefficients are defined by using BCFT techniques thanks to
 the folding trick, which turns the conformal defect into the boundary.
With this definition, exact computations are demonstrated to describe
 reflection/transmission process for a class of conformal defects. 
We also compute the boundary entropy based on the boundary state.
\end{abstract}

\end{titlepage}

\tableofcontents

\setcounter{footnote}{0}


\section{Introduction}\label{sec:intro}
A wide range of physicists -- cosmologists, condensed matter physicists,
and particle physicists -- have been attracted by anomalous scaling
behavior of matter caused by critical phenomena. 
Studying critical phenomena with conformal defects is of great
interest, because most of realistic situations inevitably contain impurities.
A powerful method for studying critical phenomena with
conformal defects is boundary conformal field theory (BCFT).
There are many applications of BCFT especially to one-dimensional
quantum systems with impurities,
e.g., the Heisenberg spin chain, the Kondo model, and so on.
See~\cite{Affleck:2008LH} for a review along this direction.
However, it has not been completely understood how BCFT describes the reflection/transmission at conformal defects.
For this purpose, Quella, Runkel, and Watts proposed the
reflection/transmission coefficient to characterize the transport
phenomena at the conformal defect~\cite{Quella:2006de}. 
Their proposal is quite natural and generic in the sense of
CFT because their coefficients are based on the gluing condition for the energy-momentum tensor.
However, it is not obvious how the proposed coefficient is related to 
transport coefficients used in other contexts, such as quantum wire
junctions and experiments.
Our goals are to further investigate the meaning of the proposed
reflection/transmission coefficient and to obtain a more detailed
description of the reflection/transmission process.


In this paper, we define the reflection/transmission coefficient for
conserved currents, as a natural generalization of 
that proposed in~\cite{Quella:2006de} and also in~\cite{Bachas:2001vj}.
Our definition involves current algebras and boundary states, which characterize boundary conditions of fields at conformal defects. 
We demonstrate exact computations for two systems: the system having
permutation boundary conditions and the system partially breaking the
$SU(2)_{k_1}\times SU(2)_{k_2}$ symmetry into the $SU(2)_{k_1+k_2}$.
Our definition also reveals which current penetrates the conformal
defects as well as how much it does.
In addition, we compute the boundary entropies 
to identify the amount of information carried by the boundary. 
In general, it is difficult to distinguish the contributions of the
boundary and the bulk CFTs to the entropy. 
In our analysis, since the boundary state is explicitly constructed, we
can separate them more efficiently, and obtain results consistent with
previous works.


\section{Reflection and Transmission coefficients}\label{sec:Transmission}
We shall briefly review the reflection/transmission
coefficient proposed in~\cite{Quella:2006de} and give a more detailed
meaning to that.
That is to say, we 
claim that the proposed reflection/transmission coefficient 
corresponds to the energy transport.
Besides, by generalizing their proposal, we define the
reflection/transmission coefficient for a conserved current with
conformal weight $h=1$.

\subsection{Conformal defect and the junction}
\label{sec:model}

\begin{figure}[t]
 \begin{center}
  \includegraphics[width=35em]{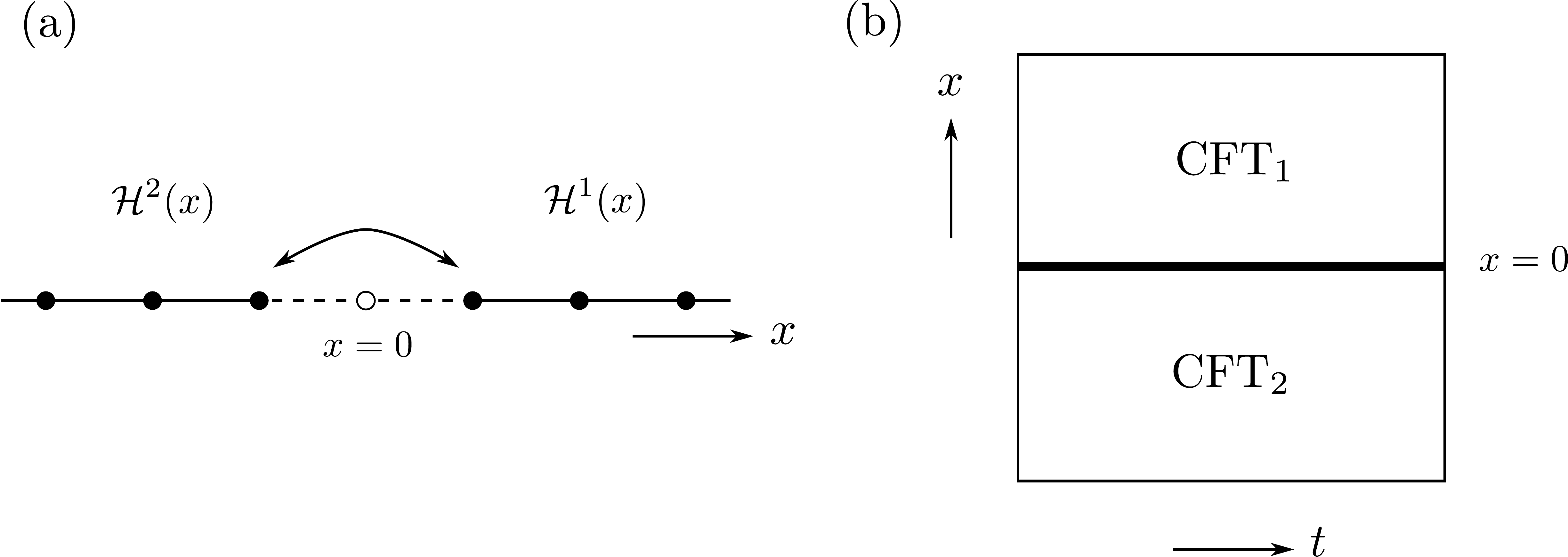}
 \end{center}
 \caption{From the {\it impurity} to the {\it defect}. (a) Two
 one-dimensional systems are connected through the impurity at
 $x=0$. (b) Adding the time direction and taking the continuum limit,
 that system is mapped into the two-dimensional system with the defect
 along the line $x=0$.}
 \label{fig:SpinChain}
\end{figure}

We consider two one-dimensional quantum systems connected by a junction, which
can be considered as an impurity interacting with the bulk.
Let us assume that the first system is in the positive domain $x>0$, the
second is in the negative $x<0$, and they are connected at the
origin as depicted in Fig.~\ref{fig:SpinChain}(a).  
If these systems obey symmetry algebras $\cA_i$,
the Hamiltonian density for each domain is obtained by Sugawara
construction at the conformal fixed point\footnote{Here we omit the
anti-holomorphic part.}
\begin{align}
 \mathcal{H}^1(x)
 & =
  \frac{1}{2\pi ( k_1 + h_1^{\vee} )} d_{A B}^1 \, J^{1,A}(x) \, J^{1,B}(x)
 \, , \quad (x > 0) 
 \label{Ham_den01} \\
 \mathcal{H}^2(x)
 & =
  \frac{1}{2\pi ( k_2 + h_2^\vee )} d_{A B}^2 \, J^{2,A}(x) \, J^{2,B}(x)
 \, , \quad (x < 0) 
 \label{Ham_den02} 
\end{align}
where $d^i_{A B}$ is the inverse of the Cartan--Killing form 
and $h_i^\vee$ is the dual Coxeter number of the algebra $\cA_i$.
The current $J^{i,A}$ takes value in the Lie algebra $\cA_i$ and
the index $A$ runs over 
$A = 1, \cdots,\mathrm{dim}\,\cA_i$.
In general, 
$\cA_1$ and $\cA_2$ can be different algebras.
The Fourier modes of $J^{i,A}$ satisfy the Kac--Moody algebra $\wh{\cA}_i$:
\begin{align}
 [j^{i,A}_m, j^{i,B}_n] 
 & = (f^i)^{A B}_{~~~C}~ j^{i,C}_{n+m} 
 + k_i \, m \, d^{i,A B} \,\delta_{m+n,0}
 \, ,
\end{align}
where $f^i$ is the structure constant of $\cA_i$ and $k_i$ is the level
of $\wh{\cA}_i$.
Especially for the $SU(2)$ theory,
this level corresponds to the electron spin as $k = 2s$ for the
multi-critical spin chain~\cite{Affleck:1985wb,Affleck:1987ch} and to
the number of channels for the Kondo
model~\cite{Affleck:1990zd,Affleck:1990by,Affleck:1990iv,Affleck:1995ge}.
Note that the anti-holomorphic parts satisfy the same Kac--Moody algebras.

In general, the impurity breaks the symmetry of the bulk theory,
and couples to a common subalgebra $\cC$ of $\cA_1$ and $\cA_2$.
One possibility for the interaction term is 
\begin{equation}
 \mathcal{H}^{\rm int}(x)
  =
  \delta(x) \, d_{ab} (\lambda_1J^{1,a} + \lambda_2J^{2,a}) S^b
  \, ,
  \label{eq:ImpurityInt}
\end{equation}
where $\lambda_i$ are the coupling constants and $J^{i,a}$ takes value in
the subalgebra $\cC$.
Here, $S^a$ stands for the impurity {\it spin} and $d^{ab}$ is the
Cartan--Killing form for $\cC$. 
For this kind of interaction, as well discussed in the Kondo problem, we
can complete the square by shifting the current $\cJ^{i,a} = J^{i,a} +
2 \pi S^a \delta(x)$ when the coupling constant takes the critical value.
Then we obtain a quadratic Hamiltonian again.
This observation indicates the existence of a non-trivial conformal
fixed point at low energy with the impurity spin absorbed. 
We remark that although \eqref{eq:ImpurityInt} is written in terms of the currents, there are models whose interaction terms should be written in terms of fundamental fields rather than currents, e.g., the spin
chain with a single impurity model~\cite{Eggert:1992ur}.
Even in such a case, it is expected that the conformal fixed point obtained
by the RG flow is described by the Hamiltonian \eqref{Ham_den01} and
\eqref{Ham_den02} with 
boundary conditions, which are specified in the following sections.

\begin{figure}[t]
 \begin{center}
  \includegraphics[width=35em]{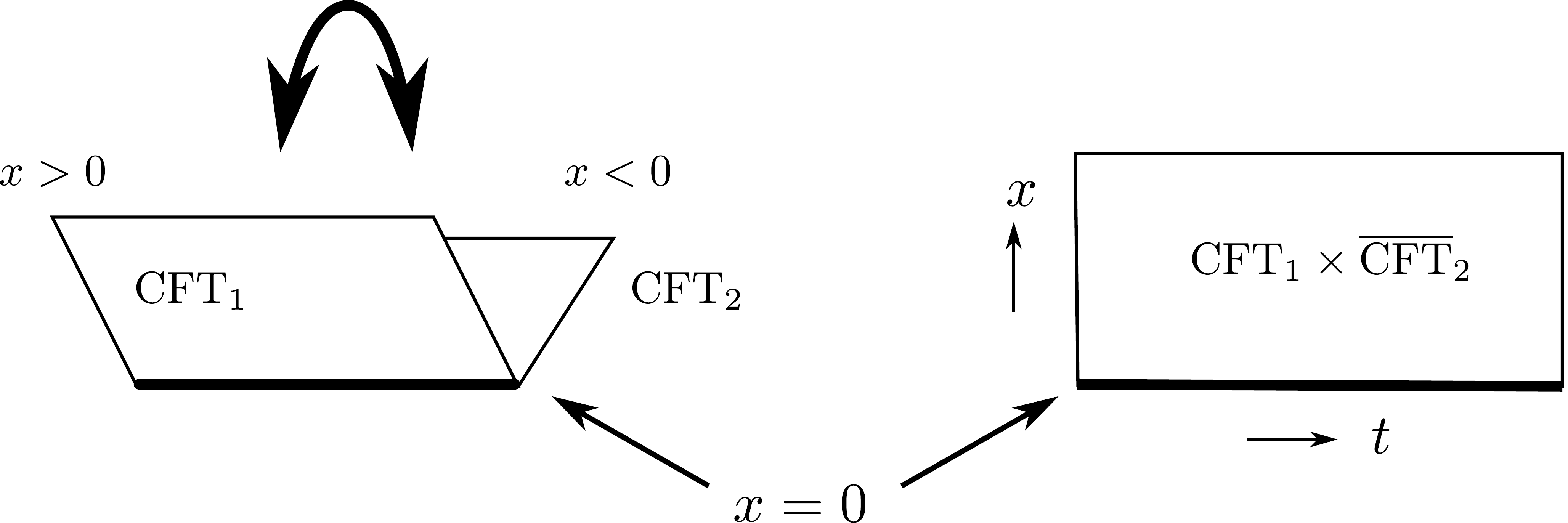}
 \end{center}
 \caption{From the {\it defect} to the {\it boundary}. By using the folding trick, a system with
 the defect is mapped into another system defined on the upper half plane
 with the boundary.}
 \label{fig:folding}
\end{figure}

Now we shall describe the above system in terms of BCFT.
Corresponding to the two quantum systems, the BCFT picture involves two CFTs: CFT$_{1}$ and CFT$_2$. 
These CFTs are defined in the upper and lower half planes respectively as depicted in Fig. \ref{fig:SpinChain}(b). 
The real axis, which divides the two CFTs, stands for the world line of the impurity, or the defect.
We can reformulate this system to obtain CFT$_1\times$$\overline{\rm
CFT}_2$ in the upper half plane thanks to the folding
trick~\cite{Wong:1994np,Oshikawa:1996ww,Oshikawa:1996dj}, as shown in Fig.~\ref{fig:folding}.
In this way, the junction of the one-dimensional quantum systems can be
mapped into a CFT boundary condition.

\subsection{Energy Reflection and Transmission}\label{sec:energy}
%
Let us then introduce the reflection/transmission coefficient to
characterize the transport phenomena at the conformal defect.
For this purpose, there are two key ingredients.
The first is a boundary state. 
The boundary state is a state of BCFT which characterizes
a boundary condition at the defect.
For example, the boundary condition for the energy-momentum tensor,
which implies 
the energy conservation at the defect, gives the
so-called Virasoro gluing condition:
\begin{align}
(L_n^{\tot}-\overline{L}_{-n}^{\tot}) \lvert B \rangle = 0 \,.
\end{align}
Here $L_{n}^{\tot}$ is the sum of Virasoro generators of CFT$_{1,2}$: 
\begin{align}
L^{\tot}_n = L_n^{1} + L_n^{2} \,.
\end{align}
The Virasoro gluing condition also ensures that the junction preserves
the conformal symmetry.%
\footnote{Beside, in string theory context, this condition
ensures that the energy flow vanishes at the open string endpoints.}

%
The second is the R-matrix,
the 2 by 2 matrix defined as~\cite{Quella:2006de}
\begin{align}
 R^{ij} 
 = 
 \frac{\langle 0 \rvert L_2^i \ol{L}_2^j \lvert B \rangle}
      {\langle 0 \rvert B \rangle}
 \, , \qquad i, j = 1, 2 \, ,
\end{align}
where $\ket{0}$ is the conformal vacuum.
Although the R-matrix has four components, it has only one degree of freedom due to the following three constraints.
The first constraint is given by the Virasoro gluing condition: 
\begin{align}
\bra{0} L^{\tot}_2 \ol{L}^{\tot}_2 \ket{B} 
 = \bra{0} L^{\tot}_2 L^{\tot}_{-2} \ket{B} 
 = \frac{c_1+c_2}2 \bra{0} B \rangle \, ,
\label{eq:ConstraintTotal}
\end{align}
where $c_{1,2}$ are the central charges for CFT$_{1,2}$, respectively.
The second and the third constraints originate from the existence of two
primary fields with respect to the total energy-momentum tensor $T^{\tot} = T^1 + T^2$ and its Hermitian conjugate: $W=c_2T_1-c_1T_2$ and $\ol{W}=c_2\ol{T}_1-c_1\ol{T}_2$.
Thus we have
\begin{align}
\bra{0} L_2^{\tot} \ol{W}_2 \ket{B} = \bra{0} W_2 \ol{L}_2^{\tot} \ket{B} = 0 \,. 
\end{align}
We remark that
these constraints show that $R^{ij}$ is symmetric:
\begin{align}
0 = \bra{0} (L_2^{\tot} \ol{W}_2 - W_2 \ol{L}_2^{\tot}) \ket{B} 
= -(c_1+c_2) \bra{0} (L_2^1\ol{L}_2^2 - L_2^2\ol{L}_2^1) \ket{B}
 \, .
\end{align}
As a result, the R-matrix is parametrized by a single real
parameter
\begin{align}
\omega_B 
 = 
 \frac2{c_1c_2(c_1+c_2)} 
 \frac{\bra{0}W_2\overline{W}_2\ket{B}}{\bra{0}B\rangle}
 \, ,
\end{align}
as
\begin{align}
R = \frac{c_1c_2}{2(c_1+c_2)} 
\left[
\left(
\begin{array}{cc}
\frac{c_1}{c_2} & 1 \\
1 & \frac{c_2}{c_1}
\end{array}
\right)
+ \omega_B
\left(
\begin{array}{cc}
1 & -1 \\
-1 & 1
\end{array}
\right)
\right] \,.
\label{eq:RforEnergy}
\end{align}

Now we give the definition of the reflection/transmission coefficient
$\cR/\cT$.
The proposal for $\cR$ and $\cT$ is~\cite{Quella:2006de}
\begin{align}
\cR &= \frac{2}{c_1+c_2} (R^{11}+R^{22}) \, ,
\\
\cT &= \frac{2}{c_1+c_2} (R^{12}+R^{21}) \, .
\label{eq:Tenergy}
\end{align}
It is easy to show that the sum is given by $\cR+\cT=1$ for any $\omega_B$,
which means the energy conservation.
Because the R-matrix is written in terms of Virasoro generators, 
we suggest that $\cR$ and $\cT$ are the reflection and transmission
coefficients for the energy transport at the defect.
We shall see that this interpretation is consistent with our definition
of the current reflection/transmission coefficient. 

\subsection{Current Reflection and Transmission}\label{sec:current}
We generalize the above construction of $\cR$ and $\cT$ to the
reflection/transmission coefficient for a conserved current with $h=1$. 
When we define the energy reflection/transmission coefficient, there are
two key ingredients: the boundary state and the
R-matrix.
In addition, there are three constraints, which originated from the
total energy-momentum tensor $T^{\rm tot}$ and the primary fields $W$
and $\ol{W}$, play an important role in counting the effective degrees
of freedom of the R-matrix. 
Here we shall take the similar process. 

We assume that CFT$_{1,2}$ have the same symmetry subalgebra $\cC$, which is preserved at the conformal defect. 
For such a defect, we choose the following current gluing condition
\begin{align}
 (j^{\tot,a}_n + \overline{j}^{\tot,a}_{-n}) \ket{B} = 0 
 \, ,
\label{eq:GluingJ}
\end{align}
where $j^{\tot,a}_n=j^{1,a}_n+j^{2,a}_n$ takes values in the Kac--Moody
algebra $\hat{\cC}$.
(Here $j_n^a$ is the Fourier mode of $J^a$.)
Notice that the signs in front of the anti-holomorphic sectors are opposite between energy and current gluing conditions due to the different parity of their conformal weights~\cite{Cardy:1989ir}. 

The straightforward generalization of the R-matrix is 
\begin{align}
R[\cC]^{ij,ab} = -\frac{\bra{0} j_{1}^{i,a} \ol{j}_1^{j,b} \ket{B}}{\bra{0} B \rangle}
 \, .
\label{eq:Rijab}
\end{align}
The extra minus sign is due to the sign difference in the gluing conditions. 
Note that since the gluing condition 
becomes rather complicated in terms of $j_n^{1,a}$ and
$j_n^{2,a}$, this R-matrix gives a non-trivial value, as we will show later.
Here we take the $n=1$ component of $j_{n}^{i,a}$ in contrast to $L_2^i$.
In fact, any positive choice of $n$ gives the same R-matrix. 
To see this fact, let us consider the following equation derived from the Virasoro gluing condition:
\begin{align}
0 = \bra{0} j_{n}^{i,a} \ol{j}_{n+1}^{j,b} (L^{\tot}_1 - \overline{L}^{\tot}_{-1}) \ket{B}
 \, .
\end{align}
Together with the commutator $[L_m^i,j_n^{i,a}]=-nj_{m+n}^{i,a}$, this leads to the recursion relation
\begin{align}
0 = n \bra{0} j_{n+1}^{i,a} \ol{j}_{n+1}^{j,b} \ket{B} - (n+1) \bra{0} {j}_{n}^{i,a} \ol{j}_{n}^{j,b} \ket{B}
 \, .
\end{align}
This relation implies that if we defined the R-matrix with
mode $n$, the matrix element $\bra{0} {j}_{n}^{i,a}
\ol{j}_{n}^{j,d} \ket{B}$ could be written in terms of \eqref{eq:Rijab}.
In addition, due to the symmetry, we have
\begin{align}
R[\cC]^{ij,ab} = - \frac{\bra{0} G\, j_{1}^{i,a}\,\, \ol{j}_1^{j,b}\, G^{-1} \ket{B}}{\bra{0} B \rangle}
 \, ,
\end{align}
where $G=\exp \{\alpha_a (j_0^{\tot,a}+\ol{j}_0^{\tot,a})\}$. 
If $\cC$ is a simple Lie algebra, this symmetry factorizes the R-matrix:
\begin{align}
R[\cC]^{ij,ab} = d^{ab} R[\cC]^{ij} 
 \, .
 \label{eq:R[C]ij}
\end{align} 
We remark that although the R-matrix can be defined with a generic algebra
rather than the common subalgebra $\cC$, the contribution only from
$\cC$ gives a non-trivial value.%
\footnote{
We can show that the reflection coefficient \eqref{eq:TforC}
gives $\cR = 1$ for $\cA_{1,2}/\cC$, and thus it implies the full
reflection process.
}

Now let us see that there are three constraints which reduce the degrees of freedom
of the R-matrix. 
The first constraint is associated with the current gluing condition. 
\eqref{eq:GluingJ} leads to
\begin{align}
\bra{0} j^{\tot,a}_1 \ol{j}^{\tot,b}_1 \ket{B} 
 = - (k_1 + k_2) d^{ab} \bra{0} B \rangle
 \, .
\end{align}
This constraint is similar to the constraint \eqref{eq:ConstraintTotal},
which is given by the Virasoro gluing condition.
To find the other two constraints, we introduce primary
fields with respect to the total current $J^{\tot} = J^1 + J^2$ (and its conjugate $\ol{J}^{\rm tot}$):
\begin{align}
K^a(z) &= k_2 J^{1,a}(z) - k_1 J^{2,a}(z) \, , \nonumber \\
\ol{K}^a(z) &= k_2 \ol{J}^{1,a}(z) - k_1 \ol{J}^{2,a}(z) \, .
\end{align}
It is easy to show that these satisfy
\begin{align}
\bra{0} K^a_1 \ol{j}^{\tot,b}_1 \ket{B} = \bra{0} j^{\tot,a}_1 \ol{K}^b_1 \ket{B} = 0
 \, .
\end{align}
Interestingly, these constraints ensure that $R[\cC]^{ij}$ is symmetric. 
In fact, we have
\begin{align}
0 = \bra{0} K^a_1 \ol{j}^{\tot,b}_1 \ket{B} - \bra{0} j^{\tot,a}_1 \ol{K}^b_1 \ket{B}
= (k_1+k_2) \bra{0} (j^{1,a}_1 \ol{j}^{2,b}_1 - j^{2,a}_1 \ol{j}^{1,b}_1) \ket{B}
 \, .
\end{align}
Because of the above three constraints, $R[\cC]^{ij}$ has only one degree of freedom. Now let us define $\omega_B[\cC]$ as
\begin{align}
d^{ab}\omega_B[\cC] = -\frac{1}{k_1k_2(k_1+k_2)} \frac{\bra{0}K^a_1\ol{K}_1^b \ket{B}}{\bra{0}B\rangle}
 \, .
\end{align}
With this $\omega_B[\cC]$, the R-matrix $R[\cC]^{ij}$ is given by
\begin{align}
R[\cC] = \frac{k_1k_2}{k_1+k_2} 
\left(
\left(
\begin{array}{cc}
\frac{k_1}{k_2} & 1 \\
1 & \frac{k_2}{k_1}
\end{array}
\right)
+ \omega_B [\cC] 
\left(
\begin{array}{cc}
1 & -1 \\
-1 & 1
\end{array}
\right)
\right)
 \, .
\label{eq:RforCurrent}
\end{align}
Obviously, this expression is similar to \eqref{eq:RforEnergy}. The
level $k_i$ plays essentially the same role to the central charge.

Now we can define the reflection and transmission coefficients: $\cR[\cC]$ and $\cT[\cC]$. 
\begin{align}
\cR[\cC] &= \frac1{k_1+k_2} (R^{11}+R^{22})
= \frac1{(k_1+k_2)^2} \left( (k_1^2+k_2^2) + 2k_1 k_2\omega_B[\cC] \right)\,,
\\
\cT[\cC] &= \frac1{k_1+k_2} (R^{12}+R^{21})
= \frac{2k_1k_2}{(k_1+k_2)^2} (1-\omega_B[\cC])
 \, .
\label{eq:TforC}
\end{align}
From \eqref{eq:RforCurrent}, it is easy to show that $\cR+\cT=1$, which
ensures the current conservation.
We remark that the identification of $\cR/\cT$ as the reflection/transmission coefficient is available provided that both of $\cR$ and $\cT$ are nonnegative. 
Although it is unclear that the nonnegative condition holds in general, we shall see that it holds for all examples considered in the present paper. 

Before ending this section, let us comment on the case with
$\cC=su(2)$ for later use. 
In this case, the Cartan--Killing form is $d^{ab}=-\delta^{ab}/2$. 
By defining $J^{\pm}=J^1\pm iJ^2$, the R-matrix \eqref{eq:R[C]ij} can be
rewritten as
\begin{align}
 R[\cC]^{ij} = - R[\cC]^{ij,-+}
= \frac{\bra{0} j_{1}^{i,-} \ol{j}_1^{j,+} \ket{B}}{\bra{0} B \rangle}
 \, ,
\label{eq:Rsu2}
\end{align}
as well as $\omega_B[\cC]$:
\begin{align}
\omega_B[\cC] = \frac{1}{k_1k_2(k_1+k_2)} \frac{\bra{0}K^-_1\ol{K}^+_1 \ket{B}}{\bra{0}B\rangle}
 \, ,
\label{eq:OmegaBsu2}
\end{align}
where we have used $d^{-+}=-1$.

\section{Application to some models}\label{sec:application}
We evaluate reflection and transmission coefficients for conserved
currents by using the above definition. 
In Sec.~\ref{sec:Permutation}, we consider the simpler case with the
permutation boundary condition, where we know the explicit form of the
boundary conditions for currents. 
On the other hand, in Sec.~\ref{sec:SU2}, we study the case where
$SU(2)_{k_1}\times SU(2)_{k_2}$ is broken into $SU(2)_{k_1+k_2}$ thanks
to the non-trivial interaction at the boundary.

\subsection{Permutation boundary condition for a sub-symmetry}
\label{sec:Permutation}
Let us first consider the simpler example where we impose the following boundary condition:
\begin{align}
J^{1,\alpha_1}(z) = \ol{J}^{1,\alpha_1}(z)\,,~~
J^{2,\alpha_2}(z) = \ol{J}^{2,\alpha_2}(z)\,,~~
J^{1,a}(z) = \ol{J}^{2,a}(z)\,,~~
J^{2,a}(z) = \ol{J}^{1,a}(z)\,,~~
\end{align}
where $\alpha_{1,2}$ and $a$ stand for the labels for $\cA_{1,2}/\cC$ and $\cC$ respectively. 
To be consistent with the boundary condition, we have to impose $k_1=k_2\equiv k_c$. 
In this example, degrees of freedom associated with $\cC$ completely
penetrate the defect, while the others are completely reflected. 
This observation suggests $\cT[\cC]=1$. 
Let us show this as follows.

Using the boundary condition, the off-diagonal elements of the R-matrix are
\begin{align}
\bra{0} j^{1,a}_1 \overline{j}^{2,b}_1 \ket{B} = \bra{0} j^{2,a}_1 \overline{j}^{1,b}_1 \ket{B}
 = - d^{ab} k_c  \bra{0} B \rangle 
 \, .
\end{align}
This and \eqref{eq:RforCurrent} immediately show that $\omega_B=-1$, $\cR^{11}=\cR^{22}=0$, and $\cR^{12}=\cR^{21}=k_c$. 
This proves the full transmission: $\cT[\cC]=1$. 
This result is in contrast to the energy transmission coefficient $\cT=2c/(c_1+c_2)$~\cite{Quella:2006de} where $c_{1,2}$ and $c$ are the central charges associated with $\cA_{1,2}$ and $\cC$. 
This is because only the fields associated with $\cC$ contribute to the transmission. 
In other words, we found that among the total degrees of freedom
$c_1+c_2$, $2c$ degrees of freedom completely penetrate and the others
are completely reflected. 
(The factor $2$ of $2c$ stems from the fact that both $j^{1,a}$ and $j^{2,a}$ contribute to the energy transport.)
This argument supports our identification of $\cT$ \eqref{eq:Tenergy} as
the total energy transmission coefficient.
The benefit of our current transmission is that we can
see more microscopic information about the transmission process. 

\subsection{$SU(2)_{k_1}\times SU(2)_{k_2} \to SU(2)_{k_1+k_2}$}
\label{sec:SU2}
Let us consider the more 
general case where $\wh\cA_{1,2}=su(2)_{k_{1,2}}$ and $\wh\cC=su(2)_{k_1+k_2}$.
The symmetry can be rewritten as
\begin{align}
SU(2)_{k_1} \times SU(2)_{k_2} =  \frac{SU(2)_{k_1} \times SU(2)_{k_2}}{SU(2)_{k_1+k_2}} \times SU(2)_{k_1+k_2}
 \, .
\end{align}
Hereafter, we use $G=SU(2)_{k_1} \times SU(2)_{k_2}$ and $H=SU(2)_{k_1+k_2}$. 
The $SU(2)_{k_1+k_2}$-preserving boundary states are characterized by
three parameters $(\rho_1,\rho_2,\rho)$ which run over
$2\rho_i=0,1,\cdots,k_i$ and $2\rho=0,1,\cdots,k_1+k_2$ with the
identification $(\rho_1, \rho_2, \rho) \sim (\frac{k_1}{2}-\rho_1,\frac{k_2}{2}-\rho_2,\frac{k_1+k_2}{2}-\rho)$~\cite{Quella:2002ct}:
\begin{align}
\ket{B(\rho_1,\rho_2,\rho)}
= \sum_{\mu_1+\mu_2+\mu\in\Z} 
\frac{ S^{(k_1+k_2)}_{\rho\mu}  S^{(k_1)}_{\rho_1\mu_1} S^{(k_2)}_{\rho_2\mu_2} }
{ S^{(k_1+k_2)}_{0\mu} \sqrt{ S^{(k_1)}_{0\mu_1} S^{(k_2)}_{0\mu_2}}}
\Bket{(\mu_1,\mu_2,\mu)} \otimes \Bket{\mu}
 \, ,
\label{eq:BoundaryState_rho}
\end{align}
with $2\mu_i \in \{0,1,\cdots,k_i\}$ and $2\mu \in \{0,1,\cdots,k_1+k_2\}$.
Here $\Bket{(\mu_1,\mu_2,\mu)}$ is an 
Ishibashi state for
$G/H$ and $\Bket{\mu}$ is a current Ishibashi state for $H$.
$S^{(k)}_{\rho\mu}$ is the modular S-matrix of $SU(2)_k$,~\cite{Altschuler:1989nm,Kac:1988tf}
\begin{align}
S^{(k)}_{\rho\mu} = \sqrt{\frac2{k+2}} \sin \left( \frac{\pi}{k+2} (2\rho+1)(2\mu+1) \right) \,.
\end{align}
Because $J^{\tot,a} = J^{H,a}$, the above boundary state
satisfies the current gluing condition \eqref{eq:GluingJ}. 


In order to compute the R-matrix, we have to deal with
$\bra{0}j^{i,a}_1\ol{j}^{j,b}_1\ket{B}$, whose non-trivial part is
reduced to $\omega_B[\cC]$
as shown in (\ref{eq:RforCurrent}).
Since $\ket{B}=\ket{B(\rho_{1},\rho_2,\rho)}$ is spanned by the Hilbert space
basis for $G/H \otimes H$, we first need to expand
$j^{i,a}_{-1}\ol{j}^{j,b}_{-1}\ket{0}$ or 
$K^{+}_{-1}\ol{K}^-_{-1}\ket{0}$ with them as
(\ref{eq:BoundaryState_rho}), and then identify
the highest weight vectors with respect to $G/H \otimes H$.

To begin with, the ground state $\ket{0}$ is mapped to the tensor
product of the ground states:
\begin{align}
\ket{0_G} = \ket{(0,0,0)} \otimes \ket{0_H}
 \, .
\end{align}
Notice that $\ket{0_G}$ in the left hand side is the ground state for
$G$, while $\ket{0_H}$ in the right hand side is that for $H$. 
To be more specific, let us focus on the holomorphic sector and
consider $j^{i,+}_{-1} \ket{0}$. 
There should be two independent states corresponding to $i=1,2$. The first one can be easily found,
\begin{align}
\frac1{\sqrt{k_1+k_2}}j^{\tot,+}_{-1} \ket{0_G} = \frac1{\sqrt{k_1+k_2}}\ket{(0,0,0)} \otimes j^{H,+}_{-1} \ket{0_H} \equiv \ket{w_1}
 \, .
\end{align}
Here we normalized the state: $||\ket{w_1}||^2=1$. 
Another state that is orthogonal to $\ket{w_1}$ is
\begin{align}
\ket{w_2} \equiv \frac{1}{\sqrt{k_1k_2(k_1+k_2)}}K_{-1}^+ \ket{0_G}
 \, .
\end{align}
Due to the commutation relation $[K_m^a,j^{\tot,b}_n]=f^{ab}_{~~c}K_{m+n}^c$, we find that $\ket{w_2}$ is the current primary state with respect to $H$. 
Because $\ket{w_2}$ is killed when $j^{\tot,-}_0$ acts three times,
it belongs to a spin-$1$ representation. 
Therefore, the $H$-part of $\ket{w_2}$ is determined:
\begin{align}
\ket{w_2} = \ket{w_{G/H}} \otimes \ket{1_H}
 \, .
\end{align}
In order to find $\ket{w_{G/H}}$, we shall investigate $L^{G/H}_1$. 
Because $\ket{1_H}$ is a 
primary state of $H$, 
\begin{align}
L^{G/H}_1 \ket{w_2} = (L^{G}_1-L^{H}_1) \ket{w_2} =  L^{G}_1\ket{w_2}~
\propto (L^1_1+L^2_1) K_{-1}^+ \ket{0_G} = 0
 \, .
\end{align}
In the last equality, we have used $[L_m^i,j_n^{i,a}]=-nj_{m+n}^{i,a}$.
Thus $\ket{w_{G/H}}$ is a 
primary state of $G/H$. 
Since the conformal weight of $\ket{w_2}$ is 1, the primary state of
$G/H$ is uniquely determined:
\begin{align}
\ket{w_2} = \ket{(0,0,1)} \otimes \ket{1_H} \,.
\end{align}
Here, we normalized the states: $||\ket{(0,0,1)}||^2=||\ket{1_H}||^2=1$.

According to~\cite{Ishibashi:1988kg}, Ishibashi states are expressed as
\begin{align}
\Bket{(0,0,0)}\otimes\Bket{0_H} &= \ket{0}\otimes\ket{\wt{0}} + \ket{w_1}\otimes\ket{\wt{Uw_1}} + \cdots
 \, ,
\nonumber\\
\Bket{(0,0,1)}\otimes\Bket{1_H} &= \ket{w_2}\otimes\ket{\wt{Uw_2}} + \cdots
 \, ,
\label{eq:Ishibashi_w}
\end{align}
where tildes stand for the anti-holomorphic parts. 
Dots involve states with higher weights and the current descendant states such as $j^{H,-}_{0}\ket{1_H}$. 
$U$ is an antiunitary operator that acts on $H$:
\begin{align}
Uj^{H,+}_nU^{-1} = j^{H,-}_n\,,~~
Uj^{H,3}_nU^{-1} = j^{H,3}_n
 \, .
\end{align}
By substituting \eqref{eq:Ishibashi_w} into \eqref{eq:BoundaryState_rho}, we obtain
\begin{align}
\ket{B(\rho_1,\rho_2,\rho)}
&=
\frac{ S^{(k_1)}_{\rho_10} S^{(k_2)}_{\rho_20 }}
{ \sqrt{ S^{(k_1)}_{00} S^{(k_2)}_{00}}}
\left(
\frac{S^{(k_1+k_2)}_{\rho 0}}{S^{(k_1+k_2)}_{00}}\Bket{(0,0,0)} \otimes \Bket{0}
+ \frac{S^{(k_1+k_2)}_{\rho 1}}{S^{(k_1+k_2)}_{01}}\Bket{(0,0,1)} \otimes \Bket{1}
+ \cdots
\right)
\nonumber \\
&= \frac{ S^{(k_1)}_{\rho_10} S^{(k_2)}_{\rho_20 }}
{ \sqrt{ S^{(k_1)}_{00} S^{(k_2)}_{00}}}\frac{S^{(k_1+k_2)}_{\rho 0}}{S^{(k_1+k_2)}_{00}}
\left(
\ket{0}
+ \ket{w_1}\otimes\ket{\wt{Uw_1}}
+ \frac{S^{(k_1+k_2)}_{00} S^{(k_1+k_2)}_{\rho 1}}{ S^{(k_1+k_2)}_{\rho 0} S^{(k_1+k_2)}_{01}}
\ket{w_2}\otimes\ket{\wt{Uw_2}} +
\cdots
\right)
 \, ,
 \label{B_state01}
\end{align}
where $\ket{0}$ in the second line stands for $\ket{0}\otimes\ket{\wt{0}}$.
The dots in the second line represent the states with higher weights as well as the descendants. 

Now we proceed to the computation of the R-matrix. 
Using \eqref{eq:OmegaBsu2}, $\omega_B$ is 
\begin{align}
\omega_B[su(2)] 
&= \frac1{k_1k_2(k_1+k_2)}  \frac{\bra{0}K^-_1 \ol{K}^+_1 \ket{B}} {\bra{0}B \rangle }
= \frac{\bra{w_2}\otimes \bra{\wt{Uw_2}} B \rangle} {\bra{0}B \rangle }
\nonumber \\
&= \frac{S^{(k_1+k_2)}_{00} S^{(k_1+k_2)}_{\rho 1}}{ S^{(k_1+k_2)}_{\rho 0} S^{(k_1+k_2)}_{01}}
 \, .
\end{align}
By substituting this into \eqref{eq:TforC}, the transmission coefficient is obtained as
\begin{align}
\cT[su(2)] 
= \frac{2k_1k_2}{(k_1+k_2)^2}
\left(
1 - \frac{S^{(k_1+k_2)}_{00} S^{(k_1+k_2)}_{\rho 1}}{ S^{(k_1+k_2)}_{\rho 0} S^{(k_1+k_2)}_{01}}
\right)
 \, .
\end{align}
Notice that $\cT$ is independent of $\rho_{1,2}$ as in the case of the
energy transmission~\cite{Quella:2006de}.
In the case with $k_1 = k_2 = 1$, corresponding to the junction
of $s=\frac{1}{2}$ Heisenberg spin chains, this transmission coefficient
only gives 0 or 1.
This is consistent with the fact that there are only full reflection and
full transmission fixed points~\cite{Eggert:1992ur}.
Another property of $\cT$ is that $\cT=0$ when $\rho=0$. 
We can explain this property as follows. 
It was found~\cite{Quella:2002ct} that for $\rho=0$ the original symmetry $G=SU(2)_{k_1}\times SU(2)_{k_2}$ is restored and the boundary state can be written as
\begin{align}
\ket{B(\rho_1,\rho_2,0)}
 =\ket{\rho_1} \otimes \ket{\rho_2}
 \, ,
\label{eq:BSrho0}
\end{align}
where $\ket{\rho_i}$ is the Cardy's boundary state~\cite{Cardy:1989ir} for CFT$_i$:
\begin{align}
\ket{\rho_i} = \sum_{\mu_i}\frac{S^{(k_i)}_{\rho_i\mu_i}}{\sqrt{S^{(k_i)}_{\rho_i0}}} \Bket{\mu_i}
 \, ,
\label{eq:CardyBS}
\end{align}
with $2\mu_i \in \{0,1,\cdots,k_i\}$.
The right hand side of \eqref{eq:BSrho0} immediately leads to the current gluing conditions for both $j^{1,a}$ and $j^{2,a}$.
Thus we obtain $R^{12}=R^{21}=0$, and hence $\cT=0$.
The full reflection, or $\cT=0$, implies that the conformal defect is decoupled from the bulk system at the critical point. 



\section{Boundary entropy}\label{sec:entropy}

In this section, we focus on the conformal defect as the impurity in the
one-dimensional quantum system.
In general, the current in the bulk theory interacts with the impurity
as \eqref{eq:ImpurityInt}, and thus this impurity contributes to the total
free energy of the system.
This means that there is also the impurity contribution to the thermodynamic
entropy.
This impurity entropy, also called boundary entropy~\cite{Affleck:1991tk},
can be detected, for example, by estimating the entanglement
entropy~\cite{Calabrese:2004eu}.  
See also~\cite{Affleck:2009JPA}.

In order to define the boundary entropy,
let us set the total length of the system $2L$ and the temperature $T$
by compactifying the time direction. 
Under this condition, the boundary entropy is defined as 
\begin{align}
 S_{\rm bdry} = \lim_{L\to\infty} \, [S(L,T)-S_0(L,T)]
 \, ,
 \label{imp_entropy01}
\end{align}
where $S_0(L,T)$ is the bulk entropy which is obtained
in the absence of the impurity.

We are especially interested in the zero temperature limit $T \to 0$.
In this case it is enough to consider the ground state contribution.
If the boundary has no interaction with the bulk, the
boundary entropy at $T=0$ must be given by the degeneracy of the
boundary ground state.
For example, if the non-interacting impurity belongs to the spin $s$
representation of $SU(2)$, we have $S_{\rm bdry}=\ln (2s+1)$.
On the other hand, when there exists an
interaction between the impurity and the bulk, that interaction leads to non-trivial
entropy in general. 

Let us compute the boundary entropy for the models considered in
Sec.~\ref{sec:SU2}.
It was shown that the boundary entropy is given by the overlap between the
boundary state and the conformal vacuum~\cite{Affleck:1991tk}:
\begin{align}
 S_{\rm bdry} = \ln \bra{0}B \rangle - \ln\bra{0} B_0 \rangle
 \, .
 \label{imp_entropy02}
\end{align}

Here $\ket{B_0}$ represents the situation in the absence of the
interaction between the impurity and the
bulk~\cite{Affleck:1991tk,Affleck:2008LH,Affleck:1995ge}:
\begin{align}
\ket{B_0} = \ket{0} \otimes \ket{0}
 \, ,
\end{align}
where $\ket{0}$'s are the Cardy's boundary states \eqref{eq:CardyBS}
with $\rho_i=0$.
Therefore, we demand that the contribution from $\ket{B_0}$ corresponds to the bulk contribution $S_0$. 
Through \eqref{eq:BSrho0}, we can rewrite $\ket{B_0}$ as
\begin{align}
\ket{B_0} = \ket{B(0,0,0)}
 \, .
\end{align}



From the expression \eqref{B_state01} the overlap between the vacuum and
the boundary state is given by
\begin{equation}
 \bra{0} B(\rho_1,\rho_2,\rho) \rangle
  =
  \frac{ S^{(k_1)}_{\rho_10} S^{(k_2)}_{\rho_20 } }
       { \sqrt{ S^{(k_1)}_{00} S^{(k_2)}_{00} } }
  \frac{ S^{(k_1+k_2)}_{\rho 0} }{ S^{(k_1+k_2)}_{00} }
  \, .
\end{equation}
With the above identification of $\ket{B_0}$, we have
\begin{align}
W_{\rm bdry} \equiv \exp \left( S_{\rm bdry} \right)
=  \frac{ S^{(k_1)}_{\rho_10} S^{(k_2)}_{\rho_20 } S^{(k_1+k_2)}_{\rho 0}}
{ S^{(k_1)}_{00} S^{(k_2)}_{00} S^{(k_1+k_2)}_{00} }
 \, .
 \label{imp_entropy03}
\end{align}
Here $W_{\rm bdry}$ stands for the degeneracy of the ground state.
Interestingly, this depends on $\rho_{1,2}$ in contrast to the reflection/transmission coefficients. 
As with the Kondo problem, we encounter non-integer degeneracies for
generic $(\rho_1,\rho_2,\rho)$, which
are indications of non-Fermi liquid behavior, and some of them may be
related to Majorana-like
excitation~\cite{Emery:1992PRB,Beri:2012PRL,Tsvelik:2013PRL,Beri:2012pe,Altland:2013vza}.
Since the boundary entropy can be detected in the entanglement
entropy~\cite{Calabrese:2004eu}, it can be a convenient criterion for
such a behavior, e.g., in numerical analysis.

In addition to the non-integer degeneracies as discussed above, we also
encounter integer ones for some $(\rho_1,\rho_2,\rho)$.
A remarkable example is $W_{\rm bdry}=2$ for $\rho_1=\rho_2=\rho=1$ with
$k_1=k_2=2$ 
that has the same symmetry as the two-channel Kondo model. 
It is known that the two-channel Kondo model usually gives a
non-integer degeneracy. 
However this example indicates that the ground state of the Kondo impurity
can have an integer degeneracy when the interaction involves the channel
current in addition to the electron spin current. 
To well understand the origins of these integer degeneracies as well as the physical meaning of $\rho$'s, further investigation is necessary.

The result obtained here provides an interesting implication also for the
spin chain models.
The situation we have discussed corresponds to the junction of 
$SU(2)$ chains with arbitrary spins $s_{1,2} = k_{1,2}/2$.
Thus the expression (\ref{imp_entropy03}) gives a quite general formula
for the impurity entropy of the spin chain junction.
It is interesting to check that the formula (\ref{imp_entropy03})
can be obtained from the spin chain models by using
another analytical method, e.g., Bethe ansatz.




\section{Summary and discussion}\label{sec:summary}
We have defined the reflection/transmission coefficient for the
conserved current at conformal defects.
The BCFT approach offers an analytic and exact method to describe the
reflection/transmission process. 
In addition, our definition provides a
microscopic description of the reflection/transmission process. 
Namely, it reveals which and how much the current penetrates the defect. 
We have also computed the boundary entropy and observed a non-integer degeneracy.


We add some comments on the Kondo problem, to which our analysis is
directly applicable.
In particular, for $k_1=k_2=2$, the model considered in
Sec.~\ref{sec:SU2} has the same symmetry as the two-channel Kondo model.
In this case the two $SU(2)_2$'s in $SU(2)_2\times SU(2)_2$ have different
meanings: the first one is for the spin and the second is for the channel. 
Hence the transmission process means exchanging of spin and channel
currents at the defect.
As in the case of Kondo impurities, it is interesting to compute
the specific heat and the resistivity. 
That computation could give further information in order to understand
the physical meaning of $(\rho_{1,2},\rho)$.


Let us comment on some possibilities beyond this work.
It is interesting to extend our analysis of $SU(2)_{k_1}\times
SU(2)_{k_2}$ into $SU(N)_{k_1}\times SU(N)_{k_2}$.
This generalization attracts attention from not only theoretical, but
also experimental point of view.
It is because such a situation could be realized experimentally with,
e.g., a quantum
dot~\cite{Cronenwett:1998Sci,Goldhaber-Gordon:1998Ntr,Schmid:1998PhysB}, or
an ultracold atomic
system~\cite{Gorshkov:2010NP,DeSalvo:2010PRL,Fukuhara:2007PRL,Taie:2010PRL,Taie:2012NP}.
Although the Kac--Moody algebra is more complicated for $N>2$, one can
use the formal expression of boundary states given in~\cite{Quella:2002ct}, 
and we can compute the R-matrix defined in (\ref{eq:Rijab}) as in the
case of the $SU(2)$ theory in principle.
The $SU(N)$ theory may give richer results corresponding to the
non-trivial fixed points since its representation theory is rather
complicated, although some of the fixed points can be unstable.
In addition, if we could take the large $N$ limit, it is interesting to
compare with the holographic methods for
BCFT~\cite{Azeyanagi:2007qj,Takayanagi:2011zk} and for the Kondo
problem~\cite{Erdmenger:2013dpa}.
Furthermore, by applying the folding trick a number of times, we can
straightforwardly generalize our analysis to the multiple junction of CFTs.
In this case, the R-matrix becomes $M\times M$ matrix with the $M$
multiplicity of the junction. 
On top of that, it turns out that the level-rank duality allows us to regard
this system as the multi-channel Kondo model. 
We are preparing a paper in this direction. 

Although we have focused on the impurity preserving the
$SU(2)$ symmetry, we can also consider the situation where $SU(2)$ is
partly broken to $U(1)$.
Such a situation could be applicable to spin transport, which is
driven by the spin-orbit interaction.
Since the spin-orbit interaction breaks $SU(2)$ spin symmetry, the
non-$SU(2)$ symmetric, or non-magnetic impurity plays an important role
in the spin transport at the junction, especially with the Rashba effect
induced at the surface. 
In this way we expect that our transport coefficients can be
experimentally observed.

Another challenging issue is to connect critical phenomena including conformal defects to string field theory. 
String field theory derives non-trivial boundary states from its solutions through the proposed formulas~\cite{Kudrna:2012re,Kiermaier:2008qu}. 
Therefore, a new boundary state could be presented by string field theory to describe a non-trivial reflection/transmission process.
For this purpose, the level truncation technique demonstrated in~\cite{Michishita:2001qn,Kudrna:2014rya} may be helpful.
In addition, it is interesting to find the interpretation of reflection/transmission coefficient from string theory point of view.

\subsection*{Acknowledgements}

We would like to thank S.~C.~Furuya and M.~Schnabl for reading the
manuscript and giving useful comments.
We are also grateful to T.~Proch\'{a}zka for valuable discussions.
The work of TK is supported in part by Grant-in-Aid for JSPS
Fellows~(\#25-4302).
The research of MM was supported by grant GACR P201/12/G028.


\bibliographystyle{ytphys}
\bibliography{interface}

\end{document}